\documentclass[aps,onecolumn,11pt,floatfix,altaffilletter,superscriptaddress,
               preprintnumbers,tightenlines,showpacs,showkeys,notitlepage,nofootinbib]{revtex4-2}

\usepackage[colorlinks=true,citecolor=blue,linkcolor=blue]{hyperref}
\usepackage[normalem]{ulem}
\usepackage{amsmath,amssymb}
\usepackage{mathtools}
\usepackage{verbatim}
\usepackage{titlesec}               % modify title and section heading format
\usepackage{epsfig}
\usepackage{graphicx}               % Standard graphics package
\usepackage{url}
\usepackage{color}
\usepackage{multirow}
\usepackage{placeins}
\usepackage[dvipsnames]{xcolor}
\usepackage{epstopdf}
\usepackage{siunitx}
\usepackage{fontawesome}
\usepackage{enumitem}
\usepackage[capitalize]{cleveref}
\usepackage{lipsum}
\usepackage{gensymb}
\usepackage{booktabs}               % nicer looking tables
\usepackage{xspace}                 % smart space after command
\usepackage{pifont}
\usepackage{marvosym }%lightning symbol
\usepackage{tikz-feynman}
\usepackage{tikz} % to draw Feynman diagrams manually
\usetikzlibrary{shapes,arrows}
\usetikzlibrary{decorations.pathmorphing,decorations.markings}
\usetikzlibrary{snakes}
\usepackage{orcidlink}  % to add ORCID 
\allowdisplaybreaks

\usepackage{bbm}
\usepackage{slashed}

\makeatletter

\renewcommand{\p@subsection}{}
\makeatother

\titleformat*{\section}{\centering\bfseries\uppercase}
\titlelabel{\thetitle\quad}
\titleformat*{\paragraph}{\bfseries}
\titlespacing*{\paragraph}{0pt}{3.25ex plus 1ex minus .2ex}{1em}

\makeatletter
\def\l@subsubsection#1#2{}
\makeatother

\begin{document}

%=============================================================================
\title{Looping Around Neutrino Charge Radius at Ultra-Near Reactor Experiments}
% \title{Neutrino charge radius and additional one-loop radiative corrections at ultranear reactor experiments}

\author{Vedran Brdar \orcidlink{0000-0001-7027-5104}}
\affiliation{Department of Physics, Oklahoma State University, Stillwater, OK, 74078, USA}
\author{Leonardo J. Ferreira Leite \orcidlink{0000-0002-4574-9336}}
\affiliation{Instituto de Física Gleb Wataghin UNICAMP,
Rua Sérgio Buarque de Holanda, Campinas, SP, Brazil}
\author{George A. Parker \orcidlink{0009-0000-1836-8696}}
\affiliation{PRISMA+ Cluster of Excellence and Institut f{\"u}r Physik,
Johannes Gutenberg-Universit{\"a}t Mainz,
55099 Mainz, Germany}
\author{Xun-Jie Xu \orcidlink{0000-0003-3181-1386}}
\affiliation{Institute of High Energy Physics, Chinese Academy of Sciences, Beijing 100049, China}
%=============================================================================

\begin{abstract}
We scrutinize the potential of upcoming ultra-near reactor neutrino experiments to detect radiative corrections in the elastic neutrino-electron scattering channel, focusing on the JUNO-TAO and CLOUD detectors, which employ advanced scintillator detection technologies. Previous reactor experiments have already constrained the electron neutrino charge radius, which is a neutrino property associated with a certain subset of the total radiative corrections, and have achieved limits that are only about an order of magnitude away from the Standard Model prediction. Our study demonstrates that JUNO-TAO and CLOUD could discover the neutrino charge radius in the near future, considering the established treatment of the charge radius. However, we show that it is necessary to go beyond this standard treatment. By including the complete set of one-loop level radiative corrections, we find a partial cancellation with the charge radius effect, reducing the experimental sensitivity to this quantity. 
Nevertheless, JUNO-TAO and CLOUD still have the potential to achieve a $5\sigma$ discovery but over longer timescales within a reasonable operational timeframe.

\end{abstract}

\maketitle

%=============================================================================
\section{Introduction}
\label{sec:intro}
%=============================================================================
\noindent
In 1956, Cowan and Reines made the first direct experimental observation of neutrinos by using a nuclear reactor at the Savannah River Site as a source of antineutrinos \cite{Cowan:1956rrn,Reines:1953pu}. More than half a century later, this type of experiment was utilized to discover the reactor mixing angle $\theta_{13}$ \cite{DayaBay:2012fng,RENO:2012mkc,DoubleChooz:2012gmf}, which is, to date, the most precisely measured mixing angle in the lepton sector. Reactor neutrino experiments still have a lot to offer to the oscillation program; in particular, JUNO \cite{JUNO:2015zny} is anticipated to conclusively measure neutrino mass ordering.

Elements of the leptonic mixing matrix and neutrino mass squared differences are not the only parameters that can be probed at reactor experiments. Specifically, electromagnetic neutrino properties can also be studied \cite{Giunti:2008ve,Studenikin:2019ggv}. Since neutrinos are not charged particles, there are no electromagnetic interactions involving neutrinos at tree level; however, this changes at the quantum level, where magnetic, electric, and anapole moments appear, along with the neutrino charge radius. The first two are proportional to neutrino mass and hence negligible in the Standard Model (SM) framework, while the anapole moment and neutrino charge radius do not vanish for $m_\nu\to 0$ and hence are appealing targets for being tested at neutrino experiments. In the massless neutrino limit, the anapole moment is equal to the neutrino charge radius, up to an $\mathcal{O}(1)$ factor \cite{Cabral-Rosetti:2002zyl}. The strongest reported reactor neutrino constraints on the neutrino charge radius come from the Krasnoyarsk \cite{Vidyakin:1992nf} and TEXONO \cite{TEXONO:2009knm} experiments; these limits are set on the electron neutrino charge radius and are only about an order of magnitude larger than the SM value. This motivates the question whether near-future reactor neutrino experiments could discover the neutrino charge radius through $\bar{\nu}_e$ -- $e^-$ scattering, which will be addressed in this work. While we will focus exclusively on reactor experiments, it is worthwhile to acknowledge recent neutrino charge radius studies in connection to other types of experiments \cite{Cadeddu:2018dux,Mathur:2021trm,MammenAbraham:2023psg,Ge:2023oag,Herrera:2024ysj}, some of which focus on the muon neutrino charge radius. 

In addition to radiative corrections associated with neutrino charge radius, we will consider further radiative corrections in $\bar{\nu}_e$ -- $e^-$ scattering, presented in detail in \cite{Tomalak:2019ibg,Hill:2019xqk}; for other studies on probing next-to-leading-order effects at neutrino experiments, see \cite{Mishra:2023jlq,Brdar:2023ttb,Mishra:2023jlq,Kelly:2024tvh}. We will, in fact, demonstrate that these additional radiative corrections are highly relevant and impact our results both qualitatively and quantitatively. Regarding the qualitative impact, these contributions cannot be isolated at reactor neutrino experiments, complicating the extraction of the neutrino charge radius. Regarding the quantitative impact, they counteract the charge radius contribution to the $\bar{\nu}_e$ -- $e^-$ cross section, making discovery at reactor experiments more challenging.

The near-future reactor neutrino experiments that we employ to study the aforementioned effects are JUNO-TAO (henceforth TAO) \cite{JUNO:2020ijm,TAO} and CLOUD \cite{CLOUD1,CLOUD2}. The former will be a near detector of the JUNO experiment, located in the vicinity of the Taishan Nuclear Power Plant in China, and the latter is based on cutting-edge opaque scintillator (LiquidO) technology \cite{LiquidOpaque2021} and will be detecting neutrinos from the Chooz Nuclear Power Plant in France. 

The paper is organized as follows. In \cref{sec:Theory}, we introduce the neutrino charge radius and discuss the cross section for $\bar{\nu}_e$ -- $e^-$ scattering in the presence of radiative corrections. In \cref{sec:experiments} we introduce the reactor neutrino experiments under consideration and in \cref{sec:analysis} we discuss antineutrino fluxes and backgrounds. In \cref{sec:results} we present our analysis framework and then show the capability of considered experiments for the discovery of radiative corrections via $\bar{\nu}_e$ -- $e^-$ scattering. In \cref{sec:conclusions}, we conclude.

%%%%%%%%%%%%%%%%%%%%%%%%%%%%%%%%%%%%%%%%%%%%%%%%%%%%%%%%%%%%%%%%%%%%%%%%%%%%%%%%%%%%%%%%
%=============================================================================
\section{Theory}
\label{sec:Theory}

\subsection{Neutrino Charge Radius}
\label{sec:CR}
\noindent
In the SM, since neutrinos are neutral elementary particles, there is no tree-level coupling of neutrinos to photons. Such a coupling, however, can be generated radiatively and is generally formulated as
\begin{equation}
    \mathcal{L}_{\rm{eff}} \supset \overline{\nu} \Lambda_\mu(q) \nu A^\mu,
    \label{eqn:EffNeutinoCouplingGamma}
\end{equation}
where $\Lambda_\mu(q)$ is neutrino electromagnetic
vertex function defined in terms of four form factors \cite{broggini2012,Giunti:2014ixa}:
\begin{equation}\begin{split}
\Lambda_\mu(q) = & \mathbbm{f}_Q\left(q^2\right) \gamma_\mu 
- \mathbbm{f}_M\left(q^2\right) i \sigma_{\mu \nu} q^\nu
+ \mathbbm{f}_E\left(q^2\right) \sigma_{\mu \nu} q^\nu \gamma_5 
+ \mathbbm{f}_A\left(q^2\right)\left(q^2 \gamma_\mu-q_\mu \slashed{q}\right) \gamma_5\,.
\end{split}\end{equation}
Here, $\mathbbm{f}_Q$, $\mathbbm{f}_M$, $\mathbbm{f}_E$, and $\mathbbm{f}_A$ are neutrino form factors of the electric charge,  magnetic dipole, electric dipole, and anapole, respectively. 

\begin{figure}
	\centering
	\includegraphics[width=0.7\textwidth]{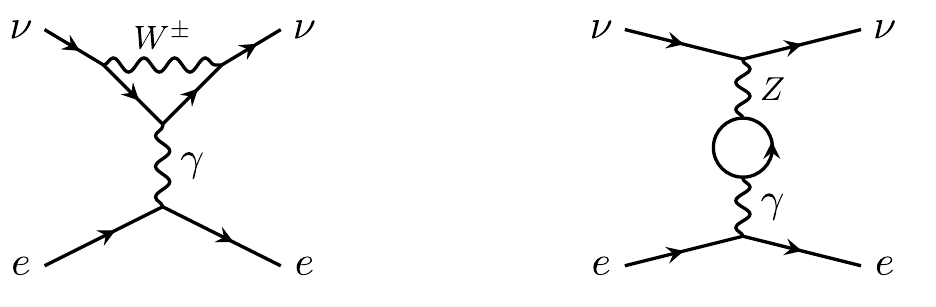}
    \caption{ Feynman diagrams that contribute to the CR flavor dependent (left) and flavor independent (right) parts.}
    \label{fig:Feyn-diag-CR}
\end{figure}

Expanding the charge form factor $\mathbbm{f}_Q$ in terms of $q^2$,
\begin{align}
\mathbbm{f}_Q\left(q^2\right)=\mathbbm{f}_Q(0)+\left.q^2 \frac{d \mathbbm{f}_Q\left(q^2\right)}{d q^2}\right|_{q^2=0}+\ldots\,,
\end{align}
one can assign physical meanings to the two terms: the first represents the electric charge of neutrinos which should be vanishing, and the second is related to the so-called neutrino charge radius (CR)~\cite{Giunti:2014ixa}. More specifically, the neutrino CR is defined as
\begin{equation}
\left.\left\langle r^2\right\rangle \equiv 6 \frac{d \mathbbm{f}_Q\left(q^2\right)}{d q^2}\right|_{q^2=0}.  
\label{eq:CRSM}
\end{equation}
In the SM, it is generated by the two diagrams shown in Fig.~\ref{fig:Feyn-diag-CR}. In the left panel, we have an explicit dependence on the charged lepton mass through the triangle loop and hence this diagram contributes to the flavor dependent part in \cref{eq:SMcr}. On the other hand, the diagram on the right is flavor independent, i.e. the associated amplitude is the same for all neutrino flavors. 

The SM prediction of the neutrino CR reads\footnote{There are different conventions involving the sign and a factor of two. See, for instance, Refs. \cite{Papavassiliou:2003rx,Vogel:1989iv,Degrassi:1989ip}.}
\begin{equation}
{\left\langle r_{\nu_{\ell}}^2\right\rangle}_{\mathrm{SM}}=-\frac{G_F}{2 \sqrt{2} \pi^2}\left[3-2 \ln \left(\frac{m_{\ell}^2}{M_W^2}\right)\right],
\label{eq:SMcr}
\end{equation}
where $G_F$ is the Fermi constant, and $m_W$ and $m_{\ell}$ are the W boson and charged lepton masses. 

Substituting specific values of $m_{\ell}$ and $m_W$ into  Eq.~\eqref{eq:SMcr}
one obtains the following CR for each lepton flavor:
\begin{equation}
{\left\langle r_{\nu_{\ell}}^2\right\rangle}_{\mathrm{SM}} =
\begin{cases}
-8.3 \times 10^{-33} \mathrm{~cm}^2, & \mbox{ if $ \ell = e $}\,,\\
-4.8 \times 10^{-33} \mathrm{~cm}^2, & \mbox{ if $ \ell = \mu $}\,,\\
-3.0 \times 10^{-33} \mathrm{~cm}^2, & \mbox{ if $ \ell = \tau $}\,.
\end{cases}
\end{equation}

Several constraints from neutrino experiments on CR already exist \cite{Giunti:2014ixa} and, for electron and muon flavor, they are only about an order of magnitude away from these SM values.

\subsection{Neutrino-electron Scattering}
\label{sec:scatter}
\noindent
Neutrino-electron scattering appears to be one of the more promising methods for probing neutrino CR experimentally. In this work, we consider reactor antineutrinos scattering off electrons, $\bar{\nu}_e e^-\rightarrow\bar{\nu}_e e^-$, for which the differential cross section at tree level reads
\begin{equation}
{\frac{d \sigma}{d T}}
=\frac{m_{e}}{4\pi}\left\{c_R^2+c_L^2\left(1-\frac{T}{E_\nu}\right)^2-c_R c_L \left(\frac{m_e T}{E_\nu^2} \right)\right\}.
\label{eq:LO}
\end{equation}
Here, $T$ is the kinetic energy of the recoil electron, $m_e$ is the electron mass,  $E_\nu$ is the  neutrino energy, and the coefficients $c_{R}$ and $c_{L}$ are given by
\begin{align}
c_{R} = 2 \sqrt{2} G_F\, s_W^2\,, &&
c_{L} = 2 \sqrt{2} G_F\left(s_W^2+\frac{1}{2}\right)\,,
\end{align}
where $s_W$ denotes the sine of the weak mixing angle. 

Equation \eqref{eq:LO} is only for tree-level calculations. Throughout this paper, we refer to it as the leading-order (LO) cross section. 
At the next-to-leading order (NLO), there are radiative corrections, with the CR being the most prominent one. 
The CR correction is often included via a shift of $\sin^2{\theta_W}$ in the literature~\cite{Giunti:2014ixa,Giunti:2015gga,Cadeddu:2018dux,Grau:1985cn}:
\begin{equation}
s_W^2 \rightarrow s_W^2\left(1+\frac{1}{3} m_W^2\left\langle r_{\nu_{\ell}}^2\right\rangle\right).
\label{eq:Giunti_shift}
\end{equation}
This shift is equivalent to  the following correction to the LO differential cross section~\cite{Tomalak:2019ibg,Hill:2019xqk}:
\begin{align}
    \delta\left({\frac{d \sigma}{d T}}
    \right)=\frac{m_e\alpha}{3} \left\langle r_{\nu_e}^2 \right\rangle \left\lvert c_R + c_L \left(1-\frac{T}{E_\nu}\right)^2 - \frac{c_R+c_L}{2} \left(\frac{m_e T}{E_\nu^2} \right) \right\rvert\,,
    \label{eq:Tomalak_shift}
\end{align}
where $\alpha$ is the fine-structure constant. 
Note that Eq.~\eqref{eq:Tomalak_shift} is an ${\cal O}(\alpha)$ correction. 
The shift of $s_W^2$ could also generate an ${\cal O}(\alpha^2)$ correction which is neglected here for a consistent truncation of radiative corrections. We have numerically checked that neglecting the ${\cal O}(\alpha^2)$ correction causes vanishingly small variation of our results.

% The radiative correction due to the CR can be readily included by adding the following correction to the LO differential cross section~\cite{Tomalak:2019ibg,Hill:2019xqk}
% \begin{align}
%     \delta\left({\frac{d \sigma}{d T}}
%     \right)=\frac{m_e\alpha}{3} \left\langle r_{\nu_e}^2 \right\rangle \left\lvert c_R + c_L \left(1-\frac{T}{E_\nu}\right)^2 - \frac{c_R+c_L}{2} \left(\frac{m_e T}{E_\nu^2} \right) \right\rvert\,,
%     \label{eq:Tomalak_shift}
% \end{align}
% where $\alpha$ is the fine-structure constant.
% This effect is equivalent to a shift of $\sin^2{\theta_W}$ in the LO cross section, 
% \begin{equation}
% s_W^2 \rightarrow s_W^2\left(1+\frac{1}{3} m_W^2\left\langle r_{\nu_{\ell}}^2\right\rangle\right),
% \label{eq:Giunti_shift}
% \end{equation}
% which is the approach often used in the literature \cite{Giunti:2014ixa,Giunti:2015gga,Cadeddu:2018dux,Grau:1985cn}.

\subsection{Radiative Corrections}
\label{sec:NLO}
\begin{figure}[h!]
    \centering
	\includegraphics[width=0.7\textwidth]{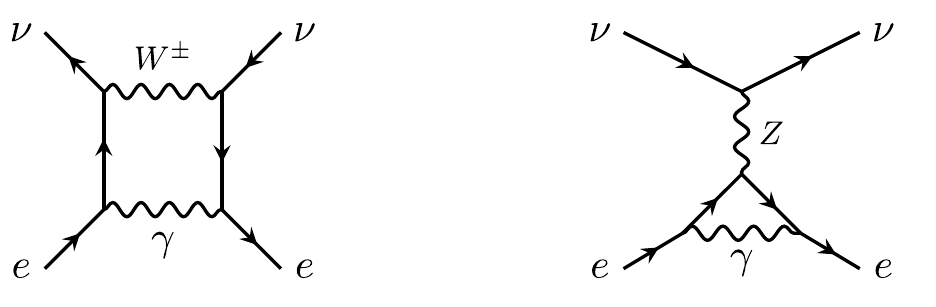}
    \caption{Examples of diagrams that are part of the remaining one-loop corrections to the antineutrino-electron scattering. On the left, we show the box-diagram and the right panel contains a representative diagram for the QED corrections.}
    \label{fig:Feyn-diag-QED}
\end{figure}

\noindent
We remark that, in addition to the CR, there are also other radiative corrections associated to the neutrino-electron scattering process. In other words, not all Feynman diagrams at the one-loop level for neutrino-electron scattering can be accounted for by an effective neutrino-neutrino-photon vertex. Specifically, besides the penguin-type diagrams and $\gamma-Z$ mixing that give rise to the CR (see \cref{fig:Feyn-diag-CR}), the process also receives radiative corrections coming from box diagrams\footnote{It can be argued that the box diagram can be considered part of the CR calculation since it provides a counter-term that cancels the gauge dependency of the other CR diagrams \cite{Lee:1977tib,Degrassi:1989ip,Papavassiliou:1989zd,Bernabeu:2000hf,Fujikawa:2003ww,Bernabeu:2004jr}. Here, however, we do not adopt this definition and consider as CR only the diagrams that contribute directly to the $\Lambda_\nu$.}, QED vertex corrections, soft-photon emission from the electron line, etc. Examples of these diagrams are shown in \cref{fig:Feyn-diag-QED}.

Ref.~\cite{Tomalak:2019ibg} consistently takes into account all relevant radiative corrections (theoretical uncertainty at the level of 0.1\%) which are, conveniently, also presented in a publicly available \texttt{Mathematica} notebook, which is employed in this work.

\begin{figure}[h!]
    \centering
	\includegraphics[width=0.7\textwidth]{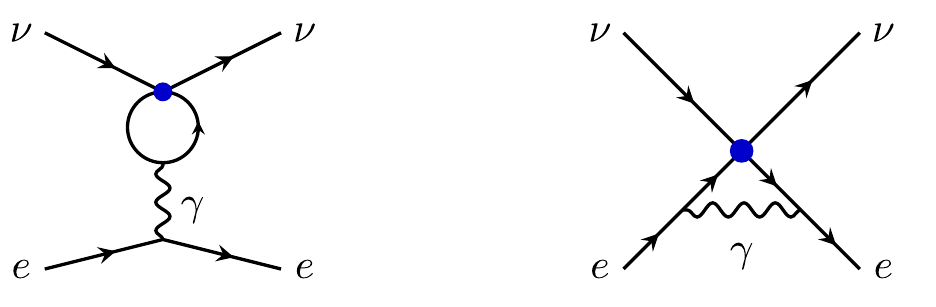}
    \caption{Representative diagrams in the effective field theory generated when the heavy degrees of freedom are integrated out. The left diagram arises from the charge radius diagrams in \cref{fig:Feyn-diag-CR}, and the right one arises from the corrections shown in \cref{fig:Feyn-diag-QED}.}
    \label{fig:Feyn-diag-CR-EFT}
\end{figure}

When treating effective operators such as the one in \cref{eqn:EffNeutinoCouplingGamma}, it is necessary to look through the lens of an effective theory obtained by integrating out the heavy degrees of freedom such as $t,W,Z$ and $h$. By performing a renormalization group evolution of relevant couplings down to the low energy scales and then matching SM onto the Low Energy Effective Theory (LEFT), diagrams such as those shown in \cref{fig:Feyn-diag-CR-EFT} appear. Here, for illustration, we show one representative diagram for CR on the left and one for QED vertex corrections on the right.

The full calculation at one-loop order, as done in \cite{Hill:2019xqk,Tomalak:2019ibg} using the effective field theory, yields the following values of the coefficients $c_L$ and $c_R$ (at renormalization scale  $\mu=2$ GeV) for electron antineutrinos scattering off electrons
\begin{align}
    &c_L = 2.39818\times 10^{-5} \text{ GeV}^{-2}\,, &&
    &c_R = 0.76911\times 10^{-5} \text{ GeV}^{-2}\,.
\end{align}

The other input parameters such as $\alpha$, $\alpha_s$, $G_F$, and $s_W^2$, are also determined at $\mu = 2$ GeV scale.\footnote{Numerical values can be found in \cite{Hill:2019xqk,Tomalak:2019ibg}.} Given that, we are armed with all input parameters necessary for calculating the cross section featuring both tree-level and radiative contributions.

%%%%%%%%%%%%%%%%%%%%%%%%%%%%%%%%%%%%%%%%%%%%%%%%%%%%%%%%%%%%%%%%%%%%%%%%%%%%%%%%%%%%%%%%%%
\begin{figure}[h!]
    \centering
    \includegraphics[width=.75\linewidth]{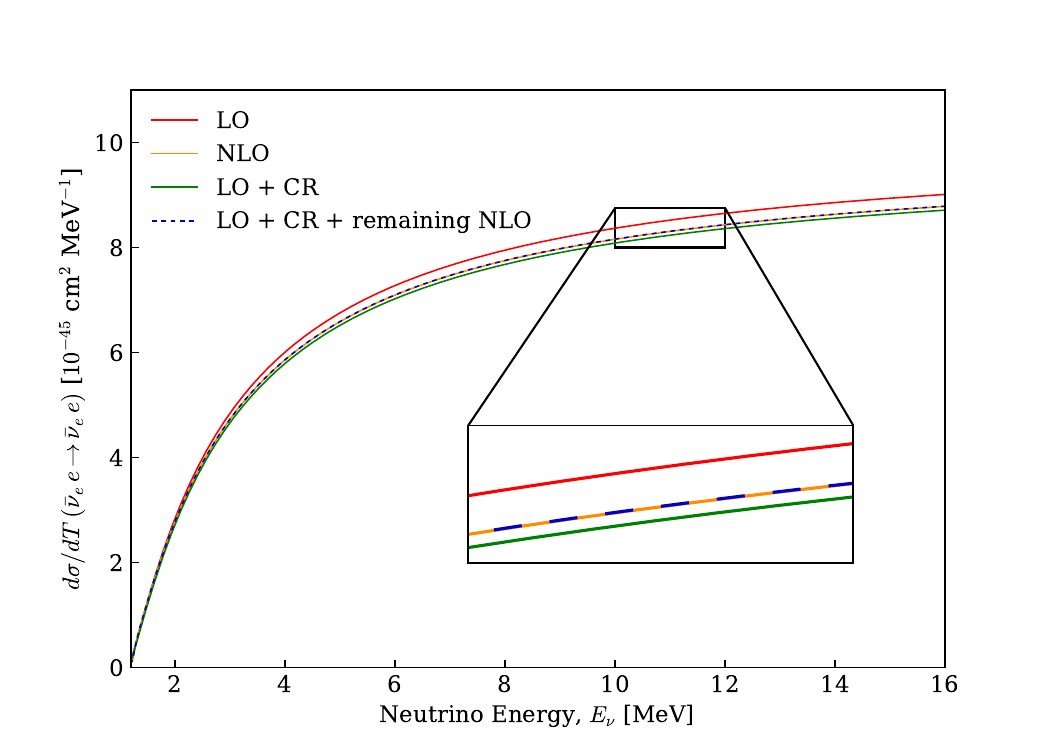}
    \caption{Differential cross section for $\bar{\nu}_e - e^{-}$ elastic scattering process as a function of the neutrino energy and for fixed electron recoil energy ($T=1$ MeV). The solid lines represent the LO cross section from \cref{eq:LO}, the full NLO cross section from \cite{Tomalak:2019ibg} and the LO with the CR contribution (LO + CR) from \cref{eq:Tomalak_shift}. The dashed line illustrates our consistency check: adding up LO + CR ($s_W^2$ shift) + ``remaining NLO'' (NLO excluding the CR) matches the NLO result from \cite{Tomalak:2019ibg}. }
    \label{fig:electron_xsec}
\end{figure}
%%%%%%%%%%%%%%%%%%%%%%%%%%%%%%%%%%%%%%%%%%%%%%%%%%%%%%%%%%%%%%%%%%%%%%%%%%%%%%%%%%%%%%%%%%

In \cref{fig:electron_xsec}, for the antineutrino-electron scattering, we show the differential cross section at next-to-leading-order (NLO) as a function of the neutrino energy (orange line). We also show the LO cross section (red line) and the cross section including only CR corrections (green line) through the shift in $s_W^2$ as discussed in \cref{eq:Giunti_shift}. 
In dashed blue, we also present  a consistency check where non-CR contributions at NLO are separately included and combined with the shift in $s_W^2$. As can be seen in \cref{fig:electron_xsec}, this curve agrees excellently with the full NLO calculation (orange), meaning that all corrections are under control and we are able to study CR-related and CR-unrelated effects separately.

We highlight that the CR has the effect of decreasing the LO cross section, while the remaining contributions to the NLO calculation have the opposite effect; this can be seen in \cref{fig:electron_xsec} by observing that the orange line is above the green one. This means that if one considers only the CR contribution without including non-CR one-loop contributions to the cross section,
the effects of radiative corrections would be stronger than the actual NLO effects. 

Since both CR and other NLO corrections can be absorbed into $c_L$ and $c_R$ parameters, we argue that it is not possible, in practical experiments, to separate CR from the remaining NLO corrections. Despite that, theoretically, by assuming a prior knowledge on the contributions of the remaining NLO corrections, one can extract CR in a model-dependent way. 
Given the above, we view it as necessary to perform a consistent calculation including the full set of NLO corrections and present discovery prospects for such a scenario instead of the one due to CR only.

%%%%%%%%%%%%%%%%%%%%%%%%%%%%%%%%%%%%%%%%%%%%%%%%%%%%%%%%%%%%%%%%%%%%%%%%%%%%%%%%%%%%%%%%
%=============================================================================
\section{Ultra-Near Reactor Neutrino Detectors}
\label{sec:experiments}
\noindent
Having in mind a few key factors such as detector fiducial mass, reactor-detector distance and reactor thermal power, we choose to focus on two forthcoming neutrino detectors within $50$ m from commercial reactor cores, TAO and CLOUD. 

\subsection{TAO}
\noindent
The JUNO (Jiangmen Underground Neutrino Observatory) experiment \cite{JUNO:2015zny, JUNO:2021vlw} is a next-generation reactor neutrino experiment, which aims to identify the neutrino mass ordering, and will begin taking data in 2025.
The JUNO detector is equidistant from two nuclear power plants: Yangjian, which has six
2.9 GW reactor cores, and Taishan, which has two 4.6 GW cores. Therefore, JUNO will (primarily) receive electron antineutrinos from reactors with a total thermal power of 26.6 GW \cite{JUNO:2021vlw}.

As this experimental program progresses, there is a simultaneous effort at the Taishan site to construct TAO (Taishan Antineutrino Observatory) \cite{JUNO:2020ijm, TAO}, a near-detector 44 m from the Taishan 1 core \cite{JUNO_neutrino24}. TAO is intended to characterise the JUNO reactor flux for the mass ordering measurement, and more generally benchmark the reactor antineutrino spectrum. 

TAO will have 2.8 ton of gadolinium-loaded liquid scintillator at $-50^{\circ}$C and use novel silicon photomultipliers (SiPMs). It will be located 9.6 m underground, where the muon and cosmogenic neutron rate are estimated to be $\sim 1/3$ of those at the surface \cite{JUNO:2020ijm}.

\subsection{CLOUD}
\noindent
On the site of the Double Chooz experiment \cite{DoubleChooz:2012gmf} there is a new next-to-next generation reactor neutrino experimental program called Super-Chooz. As a pathfinder to this ambitious project, the CLOUD (Chooz LiquidO Ultranear Detector) experiment has been proposed \cite{CLOUD1, CLOUD2}, based on cutting-edge opaque scintillator (LiquidO) technology \cite{LiquidOpaque2021}. CLOUD will have a mass of approximately 8 tons and will be positioned $\sim 35$ m from the Chooz B2 core, which produces 4.25 GW of thermal power. 

LiquidO \cite{LiquidOpaque2021} involves a lattice of wavelength-shifting fibres which is submerged in opaque liquid scintillator. Due to the short scattering length of the medium, the fibres will be able to detect scintillation photons local to their production point, giving unparalleled and fine-grained event topologies for particle identification. For the first time, this technology will allow accurate and reliable discrimination of $e^{-}$, $e^{+}$, and $\gamma$ events. Furthermore, through-going muons will produce track-like charge depositions in the detector, allowing an excellent active veto around the track to stop the most dangerous cosmogenic-induced backgrounds \cite{Conrad:2004gw}. 

However, CLOUD will be practically at the surface with only 3 meter water equivalent (mwe) overburden \cite{CLOUD1, CLOUD2}. Therefore, it is conceivable that the sheer amount of cosmogenic events could overwhelm the active veto capabilities, and reduce the efficacy of these novel handles for background rejection. In \cref{sec:analysis}, we will consider several efficiencies for background mitigation, each presented in terms of an ``effective'' overburden.

\section{Reactor fluxes and backgrounds}
\label{sec:analysis}

\subsection{Reactor Fluxes}
\label{subsec:fluxes}
\noindent
In this analysis we only consider antineutrinos from the primary reactor core. 
In both experimental setups there is a secondary core in the nuclear power plant:
for TAO, the Taishan 2 core is $252.5$ m from Taishan 1 \cite{JUNO:2021vlw} and for CLOUD, Chooz B2 is $140$ m from Chooz B1 \cite{DoubleChooz:2006vya}. By focusing only on the primary core, we are thus making a conservative estimate of the total antineutrino flux and not considering neutrino oscillations, which are negligible at this extremely short baseline.

To generate the reactor fluxes, we follow \cite{baldoncini2015}, and construct the antineutrino spectra (for ${}^{235}$U, ${}^{238}$U, ${}^{239}$Pu, and ${}^{241}$Pu) from a $5^{\textrm{th}}$ order polynomial parametrization \cite{mueller2011}. We assume a consistent load factor of 80\%, and consider the fuel fractions shown in \cref{tab:1}. For a review and comparison of reactor antineutrino spectra, see \cite{new_cite}. In our analysis we will assume reactor flux uncertainties of 0.5\% and 1\% \cite{TAO_ICHEP24}, respectively. While in the last decade there were several predictions that differed by up to $\sim 5$\%, more recent flux calculations have led to reactor antineutrino anomaly fading away \cite{Giunti:2021kab,Berryman:2021yan}. However, to reach the reactor flux uncertainties of 0.5\% and 1\%, we will rely on data-driven methods for flux determination led by PROSPECT \cite{PROSPECT:2018dnc}, and joined by both TAO and CLOUD, which will measure the fluxes in-situ and further improve our knowledge of the reactor fluxes.

%%%%%%%%%%%%%%%%%%%%%%%%%%%%%%%%%%%%%%%%%%%%%%%%%%%%%%%%%%%%%%%%%%%%%%%%%%%%%%%%
\begin{table*}[h!]
    \centering
    \setlength{\tabcolsep}{12pt}
    \begin{tabular}{c c c c c}
        \toprule
         & ${}^{235}$U & ${}^{238}$U & ${}^{239}$Pu & ${}^{241}$Pu \\
        \hline
         TAO & 0.561 & 0.076 & 0.307 & 0.056 \\\hline
         CLOUD & 0.488 & 0.087 & 0.359 & 0.067 \\
         \bottomrule
    \end{tabular}
    \caption{Fuel fractions adopted from \cite{JUNO:2020ijm, DoubleChooz:2011ymz}.}
    \label{tab:1}
\end{table*}
%%%%%%%%%%%%%%%%%%%%%%%%%%%%%%%%%%%%%%%%%%%%%%%%%%%%%%%%%%%%%%%%%%%%%%%%%%%%%%%%

\subsection{Detector Configurations and Backgrounds}
\label{subsec:bkg}
\noindent
Background considerations are important in order to assess the realistic capability of TAO and CLOUD for probing NLO effects. In our analysis, we chiefly follow \cite{Conrad:2004gw} 
where it was shown that the main backgrounds come from natural radioactivity and cosmic muons which produce spallation isotopes (i.e.~cosmogenic background). 
In \cite{Conrad:2004gw}, the latter dominates over the former for a typical liquid scintillator detector with a 300 mwe overburden. 
In our work, since both detectors have significantly less overburden, the cosmogenic background is higher than in the scenario considered in \cite{Conrad:2004gw}. Fortunately, the detector-reactor distance is also much shorter in our work compared to \cite{Conrad:2004gw}. The resulting higher neutrino flux compensates the loss caused by the cosmogenic background. This also implies that in our work, the backgrounds due to natural radioactivity would be more subdominant. 

%%%%%%%%%%%%%%%%%%%%%%%%%%%%%%%%%%%%%%%%%%%%%%%%%%%%%%%%%%%%%%%%%%%%%%%%%%%%%%%%%%%%%%%%
\begin{figure}[h!]
    \centering
    \includegraphics[width=1.\textwidth]{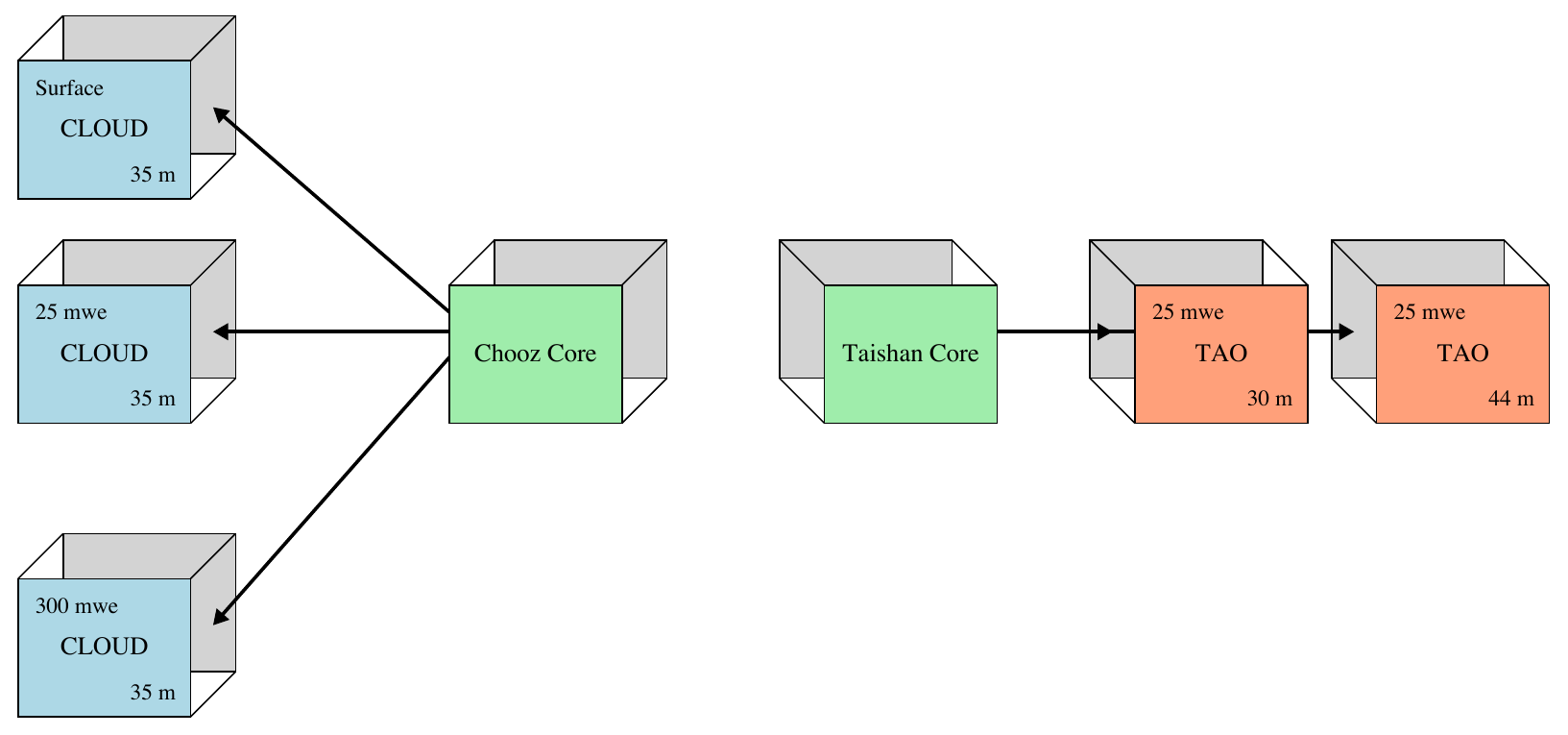}
    \caption{Schematic (not to scale) illustration of the five experimental setups considered in this work. The CLOUD detector is considered with three background scenarios, corresponding to the efficacy of the novel LiquidO technology for background rejection. The TAO detector has two possible setups with different distances from the reactor core, due to the evolving experimental design.}
    \label{fig:sketch}
\end{figure}
%%%%%%%%%%%%%%%%%%%%%%%%%%%%%%%%%%%%%%%%%%%%%%%%%%%%%%%%%%%%%%%%%%%%%%%%%%%%%%%%%%%%%%%%

For the TAO detector, we conservatively estimate that the shielding corresponds to $\sim 25$ mwe. In order to take that into account, we adopt cosmogenic backgrounds at 300 mwe from \cite{Conrad:2004gw}, and scale them to $\sim 25$ mwe using Table VII in the same work.

In light of the uncertainty regarding the efficacy of the novel technology in the CLOUD detector, we consider three distinct scenarios for its background reduction capabilities. First, for a pessimistic scenario, where the background rejection is not effective at the surface, we apply the aggressive ``surface-level'' backgrounds. For that, we use the fact that the cosmic muon rate at the sea-level  is roughly 3 times larger compared to the scenario with 25 mwe for which the muon rate reads $88.3 \,\textrm{m}^{-2}\textrm{s}^{-1}$.   
Then, we also consider a moderate scenario where the background reduction is effective in such a way that remaining background rates equal to those expected if there were 25 mwe shielding. Finally, we consider the most optimistic case, where the novel LiquidO technology is extremely effective, reducing the background to that which we expect at 300 mwe. These three configurations are depicted in the left part of \cref{fig:sketch}. 

We should also stress that there are two discussed distances between TAO detector and the reactor core reported in the literature; one realization would be achieved with 30 m baseline \cite{JUNO:2020ijm,TAO} and the other one with 44 m \cite{JUNO_neutrino24}. We will therefore consider both options; they are depicted in the right part of \cref{fig:sketch}.

\section{Results}
\label{sec:results}
\begin{figure}[h!]
    \centering
    \includegraphics[width=.7\textwidth]{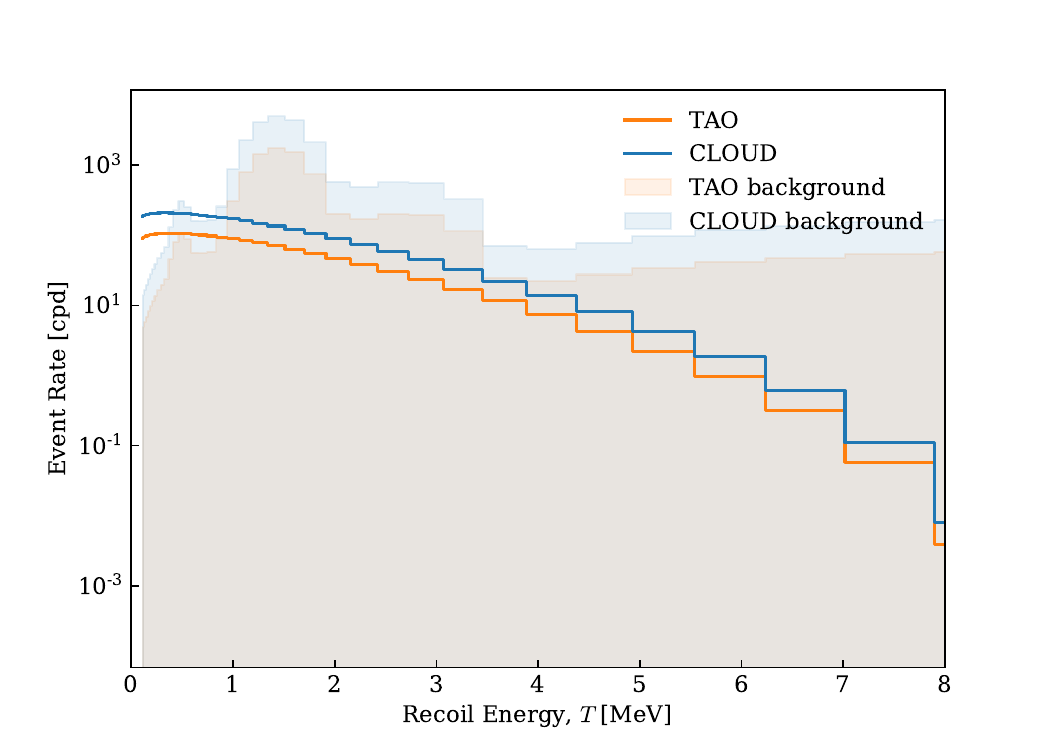}
    \caption{Comparison of LO and background event rates in units of counts per day (cpd). In this figure, we consider the moderate background scenario (25 mwe) for CLOUD, and the 30 m baseline scenario for TAO.}
    \label{fig:bkg}
\end{figure}

\noindent
Armed with the LO and NLO cross sections from \cref{sec:NLO} and reactor antineutrino fluxes from \cref{subsec:fluxes} we are able to compute the number of expected events. In order to do so, we first calculate the differential number of events with respect to the electron recoil energy:
\begin{equation}
\frac{dN}{dT}= N_{e}\Delta t \int {\frac{d \sigma\left(T,E_{\nu}\right)}{d T}}
\phi_{\bar{\nu}_e}(E_{\nu})dE_{\nu}\thinspace\,.
\label{eq:dNdT}
\end{equation}
Here, $N_e$ is the number of electrons in the detector's fiducial volume, $\Delta t$ is the data taking time, and $\phi_{\bar{\nu}_e}$ denotes the antineutrino flux. In order to obtain the total number of events in the $i$-th $T$ bin ($N_i^{\text{LO}}$ at LO and $N_i^{\text{NLO}}$ at NLO), we integrate $dN/dT$ across a given bin. Following the discussion in \cref{subsec:bkg}, we are also equipped to calculate the total background counts in $i$-th bin ($B_i$). A comparison of background event rate and LO one is shown in \cref{fig:bkg}.

%%%%%%%%%%%%%%%%%%%%%%%%%%%%%%%%%%%%%%%%%%%%%%%%%%%%%%%%%%%%%%%%%%%%%%%%%%%%%%%%%%%%%%%%
%%%%%%%%%%%%%%%%%%%%%%%%%%%%%%%%%%%%%%%%%%%%%%%%%%%%%%%%%%%%%%%%%%%%%%%%%%%%%%%%%%%%%%%%

To calculate the sensitivity to NLO effects, we carry out a binned $\chi^2$ analysis \cite{Cadeddu:2018dux} where we compare the event rates at LO and NLO
\begin{equation}
\chi^2=  \sum^{\textrm{\# bins}}_{i=1}\left(\frac{N_i^{\textrm{LO}}+ B_i-(1+\alpha) N_i^{\textrm{NLO}}-(1+\beta) B_i}{\sqrt{N_i^{\textrm{LO}}+B_i}}\right)^2 +\left(\frac{\alpha}{\sigma_\alpha}\right)^2+\left(\frac{\beta}{\sigma_\beta}\right)^2 \,.
\end{equation}

The statistical uncertainty is represented by $\sigma_i ={\sqrt{N_i^{\textrm{LO}}+B_i}}$, and $\alpha, \beta$ parameterize systematic uncertainties of the signal and background, respectively; $\alpha$ and $\beta$ are marginalized over in our $\chi^2$ analysis. We set the uncertainty of the background ($\sigma_\beta$) to 10\%, and consider two optimistic but feasible flux uncertainties of $0.5\%$ and $1\%$ (see again discussion in \cref{subsec:fluxes}). 

We choose to carry out this analysis by making a cut on the visible energy such that the interval $1-10$ MeV is included in the analysis. Below 1 MeV, the detection and event identification of elastic neutrino-electron scattering become inefficient. Above 10 MeV, the signal is small since the reactor fluxes start rapidly falling and the number of antineutrino-induced events is not competitive in that range with the large cosmogenic backgrounds.

%%%%%%%%%%%%%%%%%%%%%%%%%%%%%%%%%%%%%%%%%%%%%%%%%%%%%%%%%%%%%%%%%%%%%%%%%%%%%%%%%%%%%%%%
\begin{figure}[t]
    \centering
    \includegraphics[scale=0.6]{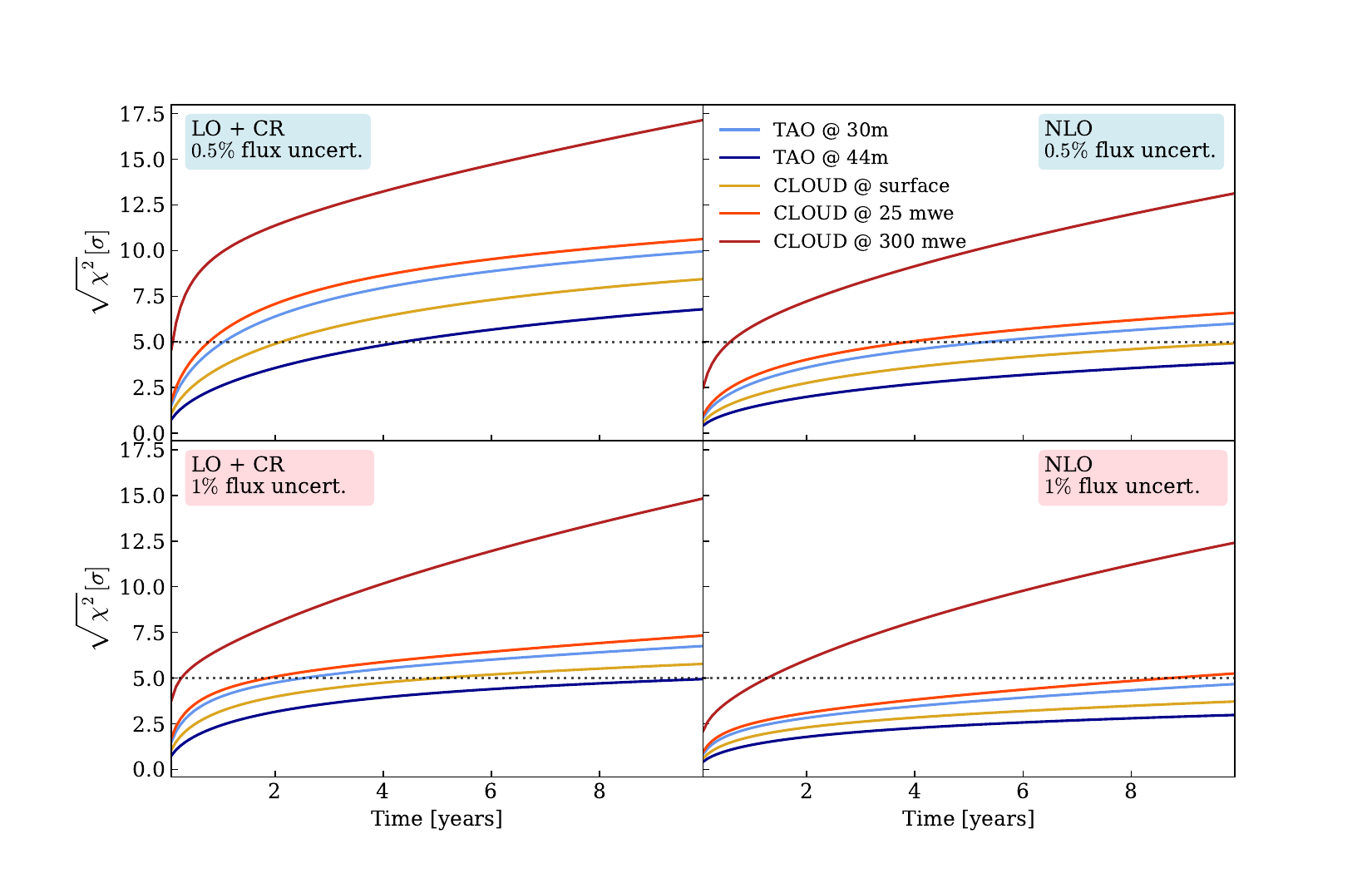}
    \caption{Sensitivity curves for data taking time ranging from 1 month to 10 years. The different lines represent the three CLOUD and two TAO detector configurations, shown for flux uncertainties of 0.5\% (top) and 1\% (bottom), and for LO with CR (left), as well as for the full set of NLO corrections (right). A black dotted line denotes 5$\sigma$ level.}
    \label{fig:results}
\end{figure}
%%%%%%%%%%%%%%%%%%%%%%%%%%%%%%%%%%%%%%%%%%%%%%%%%%%%%%%%%%%%%%%%%%%%%%%%%%%%%%%%%%%%%%%

In \cref{fig:results}, we present the results from this analysis for all five detector configurations shown in \cref{fig:sketch}. 
In the top and lower panels, we show results for 0.5\% and 1\% flux uncertainties, respectively. 
In the left panels,  we show the hypothetical sensitivity for discovering the CR, while the actual sensitivity to the full NLO corrections is shown in right panels. 
The former is obtained by assuming that only corrections associated to the CR are relevant.  
However, as we have discussed in \cref{sec:NLO}, this is not a consistent treatment. We stress again that it is not possible to measure the CR alone in neutrino-electron scattering. 

For CLOUD at the surface (gold) in the hypothetical scenario with the CR only in radiative corrections, the experiment can reach 5$\sigma$ sensitivity to the CR after about 2 or 5 years of data taking, assuming 0.5\% or 1\% flux uncertainty, respectively.
The sensitivity for the scenario including the whole set of NLO corrections is such that 5$\sigma$ is reached after 10 years (0.5\% flux uncertainty). The results are in general weaker in such case because the NLO cross section is closer to the LO one than LO+CR cross section is (see again \cref{sec:NLO}). Because of the background reduction, we overall see an improvement across all panels when considering CLOUD with 25 mwe (orange) and 300 mwe (red), respectively. As far as TAO is concerned, we find better sensitivity for the 30 m setup, simply because of the larger antineutrino fluxes. In particular, we estimate that TAO at 30 m (light blue curve) and with 0.5\% flux uncertainty would reach hypothetical discovery of the CR after 1 year, and for the real scenario where the full set of NLO corrections is included, this experiment will take 5 years and 3 months to reach 5$\sigma$. The latter amount of time still remains within reasonable operational timeframe. In conclusion, we find that both TAO and CLOUD will be capable for conclusively measuring the NLO effects in the near future.

\section{Conclusions}
\label{sec:conclusions}
\noindent
In this work, we have investigated whether radiative corrections in antineutrino-electron scattering can be observed at ultra-near reactor experiments, specifically TAO and CLOUD. We began by carefully examining the radiative corrections, noting that those related to the neutrino charge radius are not the only relevant ones. Consequently, the effects of the charge radius cannot be consistently isolated in the antineutrino-electron scattering channel. Furthermore, the remaining radiative corrections counteract the charge radius effects, bringing the next-to-leading order cross section closer to the leading order result than the cross section where only the charge radius contribution is added to the leading order. This indicates that the prospects for discovering radiative corrections are weaker than those for the hypothetical scenario including the charge radius only, as presented in \cref{fig:results}. Nevertheless, the same figure also demonstrates that both TAO and CLOUD will be capable of measuring NLO effects at $5\sigma$ within a relatively short operational timeframe. While we find the prospects for detecting higher-order corrections at ultra-near reactor neutrino experiments to be very promising, we emphasize that studying the charge radius will require the use of additional detection channels.

\begin{acknowledgments}
\noindent
We thank Stefan Schoppmann, Cloé Girard-Carillo, Alfons Weber, Hans Steiger and Michael Wurm for useful discussions. Being in the middle time zones, GAP and LJFL would like to thank the other authors of this paper for  the early mornings and late nights. GAP and LJFL are also grateful for the hospitality of CERN TH department, where this work was initially conceived.
The work of VB is supported by the United States Department of Energy Grant No. DE-SC0025477. GAP is supported by Cluster of Excellence \textit{Precision Physics, Fundamental Interactions and Structure of Matter} (PRISMA${}^{+}$ EXC 2118/1) funded by the DFG within the German Excellence strategy (Project ID 39083149). LJFL is thankful for the support of CAPES under grants No. 88887.613742/2021-00 and 88887.716533/2022-00.
The work of XJX is supported in part by the National Natural Science Foundation of China under grant No.~12141501 and also by the CAS Project for Young Scientists in Basic Research (YSBR-099). 
VB would like to thank 
the Center for Theoretical Underground Physics and Related Areas (CETUP*) and the
Institute for Underground Science at Sanford Underground Research Facility (SURF) for
providing a conducive environment during the 2024 summer workshop. This work was performed in part at Aspen Center for Physics, which is supported by National Science Foundation grant PHY-2210452. 

\end{acknowledgments}

%-----------------------------------------------------------------------------
\bibliographystyle{JHEP}
\bibliography{refs}
%-----------------------------------------------------------------------------

\end{document}